\documentclass{PoS}

\title{Implications of gamma-ray and neutrino observations on source models of ultrahigh energy cosmic rays}

\ShortTitle{Gamma-ray and Neutrino observations and source models of UHECR}

\author{\speaker{A.D. Supanitsky} \\ 
        Instituto de Tecnolog\'ias en Detecci\'on y Astropart\'iculas (CNEA, CONICET, UNSAM),
        Centro At\'omico Constituyentes, San Mart\'in, Buenos Aires, Argentina.\\
        E-mail: \email{daniel.supanitsky@iteda.cnea.gov.ar}}

\abstract{The origin and nature of the ultrahigh energy cosmic rays (UHECRs) are still unknown. 
However, great progress has been achieved in past years due to the observations performed by 
the Pierre Auger Observatory and Telescope Array. Above $10^{18}$ eV the observed energy spectrum 
presents two features: a hardening of the slope at about $10^{18.7}$ eV, which is known as 
the ankle and a suppression at approximately $10^{19.6}$ eV. The composition inferred from 
the experimental data, interpreted by using the current high energy hadronic interaction models, 
seems to be light below the ankle, showing a trend to heavier nuclei for increasing values of the 
primary energy. Current high energy hadronic interaction models, updated by using Large Hadron 
Collider data, are still subject to large systematic uncertainties, which makes difficult the 
interpretation of the experimental data in terms of composition. On the other hand, it is very well 
known that gamma rays and neutrinos are produced by UHECRs during propagation from their sources, 
as a consequence of their interactions with the radiation field present in the universe. The flux 
at Earth of these secondary particles depends on the source models of UHECRs including the chemical 
composition at injection. Therefore, both gamma-ray and neutrino observations can be used to constrain 
source models of UHECRs, including the composition in a way which is independent of the high energy 
hadronic interaction models. In this article I will review recent results obtained by using the 
latest gamma-ray and neutrino observations.
}

\FullConference{International Conference on Black Holes as Cosmic Batteries: UHECRs and Multimessenger 
        Astronomy - BHCB2018\\
		12-15 September, 2018\\
		Foz du Iguazu, Brasil}

\begin{document}

\section{Introduction}

Although the origin of the ultrahigh energy cosmic rays ($E\geq 10^{18}$ eV, UHECRs) is still an open 
problem of high-energy astrophysics, considerable progress has been reached in recent years due to the 
high quality data collected by the Pierre Auger and Telescope Array observatories. 

There are three main observables to investigate the nature of these very energetic particles: the energy
spectrum, the distribution of the arrival directions, and the chemical composition. 

The UHECR flux has been measured by both the Pierre Auger Observatory \cite{AugerSpec:17} and Telescope 
Array \cite{TASpec:15}. The flux measured by these two experiments presents a steepening known as the ankle 
at an energy of $\sim 10^{18.7}$ eV and a suppression at higher energy. These two independent measurements 
present differences. However, it has been shown that increasing the Auger energy scale in 5.2\% and 
decreasing the Telescope Array energy scale in 5.2\% (this value is within the systematic uncertainties 
of both experiments), the two measurements become compatible in the ankle region \cite{AugerTASpec:17}. 
There are still discrepancies in the suppression region.

Recently, Auger found that the cosmic ray flux presents a dipole anisotropy above $8\times 10^{18}$ eV,
detected at more than $5.2\, \sigma$ level of significance \cite{Science:17}. This result suggests an 
extragalactic origin of the cosmic rays in this energy range \cite{Science:17}. Regarding point source 
searches, we can say that at present there is no identified point source \cite{AugerPS:15}. There is a 
correlation between the cosmic ray arrival directions and the position of starburst galaxies 
\cite{AugerSBG:18}. However, the statistical significance of this correlation is at the level of 
$\sim 4\, \sigma$. Therefore, at present the astronomical objects in which the acceleration of the 
UHECRs take place are still unidentified.     

One of the parameters most sensitive to the primary mass is the atmospheric depth at which the cosmic
ray cascades reach their maximum development, $X_{\textrm{max}}$. This parameter can be obtained
by the fluorescence telescopes of Auger and Telescope Array. The composition of the UHECRs is inferred 
by using simulations of the showers, which make use of high energy hadronic interaction models (HEHIMs). 
The HEHIMs extrapolate low energy accelerator data to the highest energies. They present important 
systematic uncertainties due to the fact that the high energy hadronic interactions cannot be obtained 
from first principles. The $X_{\textrm{max}}$ measured by Auger as interpreted by using current HEHIMs 
is compatible with a light composition between $10^{18}$ eV and $\sim 10^{18.4}$ eV \cite{AugerXmax:17}. 
At this energy a transition to a heavier and mixed composition is observed. On the other hand, the 
Telescope Array measurement, interpreted by using current HEHIMs, is compatible with a light composition
up to the highest energies observed \cite{TAXmax:18}. However, the lack of statistics in the highest
energy bins do not allow rejection of the presence of elements heavier than protons and helium nuclei.

The UHECRs can interact with the low energy photons of the radiation field present in the Universe during 
propagation through the intergalactic medium. The relevant photon backgrounds are the cosmic microwave 
background (CMB) and the extragalactic background light (EBL), which in turn is composed by the infrared, 
optical, and ultraviolet backgrounds. A flux at Earth of very energetic neutrinos and gamma rays is 
expected as a result of these interactions. Neutrinos are produced by the decay of charged pions, as it 
was first pointed out by Berezinsky and Zatsepin \cite{Berezinsky:69}, and also by the decay of neutrons. 
Gamma rays are mainly produced by the decay of neutral pions. Unlike neutrinos, these very high energy 
gamma rays interact with the radiation field present in the intergalactic medium generating electromagnetic 
cascades. Electron-positron pairs, produced by the interactions of the cosmic rays during propagation, can 
also contribute to these electromagnetic cascades. The neutrino and gamma-ray fluxes at Earth, predicted 
by different models of the extragalactic cosmic ray sources, are strongly dependent on the assumptions of 
those models including the injected composition. Therefore, gamma-ray and neutrino observations can be 
very useful to constrain such models. Moreover, they can help to elucidate the presence of heavy nuclei 
at the highest energies in an independent way from the composition analyses that are based on HEHIMs.

\section{Gamma-ray and neutrino observations}
\label{GammaNuObs}

The gamma ray background that fills the intergalactic medium has recently been measured by the Fermi-LAT 
instrument \cite{FermiLAT:15}. The extragalactic gamma-ray background (EGB) is composed of the contribution 
of individual detected sources and a residual isotropic diffuse gamma-ray background (IGRB). It is believed 
that the IGBR is formed by different components, mainly undetected gamma-ray sources, i.e. which have a flux 
level smaller than the sensitivity of Fermi-LAT (see Ref.~\cite{Fornasa:15} for a review). The principal 
candidates are unresolved blazars, misaligned active galactic nuclei, starburst galaxies, and millisecond 
pulsars. Also gamma rays originated by the interactions of the UHECRs during propagation can contribute to 
the IGRB. Finally, a more uncertain contribution can come from the gamma rays originated by the decay or 
annihilation of dark matter particles and also as a result of primordial black hole evaporation.
 
\begin{figure}[!ht]
\centering
\includegraphics[width=10cm]{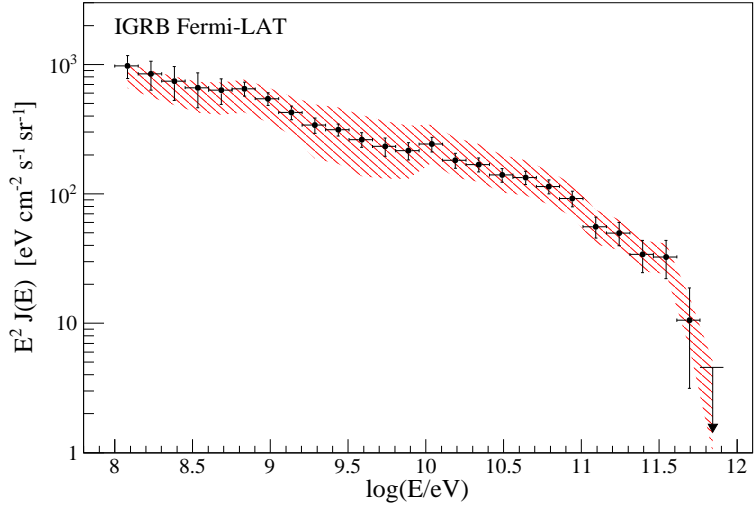}
\caption{Isotropic gamma-ray background obtained by Fermi-LAT \cite{FermiLAT:15}. The shadowed region indicates 
the systematic uncertainties due to the modeling of the galactic diffuse foreground. 
\label{Gamma}}
\end{figure}

The IGRB measured by Fermi-LAT was obtained from 50 months of data covering the energy range from 100 MeV
to 820 GeV. The events used in the analysis are such that $\left| b \right| > 20^\circ$, where $b$ is the
galactic latitude. The diffuse galactic foreground has to be removed in order to obtain the IGRB. For that 
purpose three different models were consider by the Fermi-LAT Collaboration. Figure \ref{Gamma} shows the 
IGRB obtained by Fermi-LAT by using model A to subtract the galactic foreground (see Ref.~\cite{FermiLAT:15} 
for details). A good fit of the data is obtained by considering a power law with an exponential cutoff. The 
spectral index and the cutoff energy obtained from the fit are $\sim 2.3$ and $\sim 280$ GeV, respectively.

In a subsequent analysis, the Fermi-LAT Collaboration was able to estimate the source count distribution 
($dN/dI_\gamma^{PS}$, where $I_\gamma^{PS}$ is the integral of the differential flux in a given energy interval), 
including the region below the source detection threshold, by using detailed Monte Carlo simulations and data 
\cite{AckermannPS:16}. Considering the estimated source count distribution, they could assess the integrated EGB 
flux, corresponding to the energy range from 50 GeV to 2 TeV, that originates in point sources. The obtained value 
is $I_\gamma^{PS}=2.07^{+0.40}_{-0.34}\times 10^{-9}$ cm$^{-2}$s$^{-1}$sr$^{-1}$, which corresponds to 
$86^{+16}_{-14}\%$ of the EGB flux. Making use of this result, an upper limit to the integrated gamma-ray flux 
that do not originate in point sources was calculated in Ref.~\cite{Supanitsky:16}. The values of the upper 
limit at $90\%$ and $99\%$ confidence level (CL) are given by,
\begin{equation}
I_\gamma^{UL} = \left\{ 
\begin{array}{ll}
9.345 \times 10^{-10}\ \textrm{cm}^{-2} \textrm{s}^{-1} \textrm{sr}^{-1} & 90\%\ \textrm{CL} \\
& \\
1.258 \times 10^{-9}\ \textrm{cm}^{-2} \textrm{s}^{-1} \textrm{sr}^{-1} & 99\%\ \textrm{CL}
\end{array}    \right.. 
\end{equation}

Therefore, all models of the cosmic ray sources should produce a gamma-ray flux at Earth smaller than the 
measured IGRB and also the integral of this gamma-ray flux between 50 GeV and 2 TeV should be smaller than 
$I_\gamma^{UL}$, at a given CL. 

\begin{figure}[!ht]
\centering
\includegraphics[width=10cm]{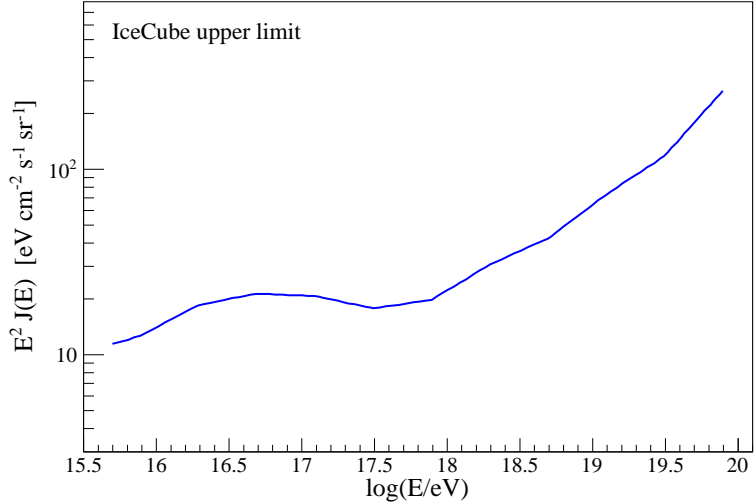}
\caption{All-flavor neutrino flux upper limit at 90 \% CL obtained by IceCube \cite{IceCube:18}. 
\label{Nu}}
\end{figure}

On the other hand, the IceCube Collaboration obtained the most restrictive differential upper limit on the 
all-flavor neutrino flux, $J_\nu^{UL}(E)$, in the energy range from $5 \times 10^{15}$ eV to $5 \times 10^{19}$ eV 
\cite{IceCube:18}. This upper limit has been obtained by using nine years of data for the analysis. In the data 
set two neutrino events with energies above $10^{15}$ eV were found. Also no neutrino event was found above 
$10^{16}$ eV. Fig.~\ref{Nu} shows the upper limit at 90 \% CL obtained by IceCube. Note that the two events 
observed increase the upper limit in the energy range below $4\times10^{17}$ eV.

Also in this case, all models of the cosmic ray sources should produce a neutrino flux at Earth smaller than 
the upper limit on the all-flavor flux obtained by IceCube.

\section{Constraints on source model of UHECRs}

As was first proposed in Ref.~\cite{Berezinsky:75}, gamma-ray observations of the IGRB were used to constrain 
proton models of UHECRs in order to predict the cosmogenic neutrino flux at Earth (see for instance 
Refs.~\cite{Ahlers:10,Berezinsky:11,Gelmini:12}). That was very important for the design of the first generation
neutrino observatories.  

As mentioned before, proton models of UHECRs are disfavored by the Auger composition results obtained by using 
current high energy hadronic interaction models to interpret the experimental data. It is very well known that 
the cosmogenic gamma-ray and neutrino fluxes are smaller in scenarios where the UHECRs are composed by heavier
nuclei. Therefore, gamma-ray and neutrino observations can be very useful to test the presence of a large 
fraction of protons in the data. In fact, in Ref.~\cite{Heinze:16} it was shown that proton models of UHECRs 
that fit the Telescope Array energy spectrum above $10^{18.2}$ eV, for which the evolution of the spatial density 
of sources with redshift $z$ is the one corresponding to the star formation rate multiplied by $(1+z)^m$, is 
rejected at 90\% CL by the IceCube upper limit. However, as was shown in 
Refs.~\cite{Supanitsky:16,Berezinsky:16,Gavish:16,Globus:17}, there are still proton models of UHECRs compatible 
with gamma-ray and neutrino observations. 

As an example of this type of study, let us consider the analysis performed in Ref.~\cite{Supanitsky:16}. Here, 
the results obtained by updating the upper limit on the all-flavor neutrino flux inferred from IceCube data (see 
Sec.~\ref{GammaNuObs}) are reported. 

Following Ref.~\cite{Supanitsky:16}, it is assumed that the UHECRs are protons injected by sources that 
are uniformly distributed in the Universe. The protons are injected following a spectrum which is given by,
\begin{equation}
\phi(E,z) = C\ S(z)\ E^{-\alpha} \exp(-E/E_{cut}),
\label{Jinj}
\end{equation}   
where $C$ is a normalization constant, $\alpha$ is the spectral index, $E_{cut}$ is the cutoff energy, and $S(z)$ 
parametrizes the evolution of the spatial density of sources with redshift $z$. The evolution function $S(z)$ is not known,
it depends on the types of sources responsible for the acceleration of the cosmic rays, which are still unknown. Here a
broken power law of $(1+z)$ is assumed,
\begin{equation}
S(z) = \left\{ 
\begin{array}{ll}
(1+z)^m & z\leq 1 \\
2^{m-n}\ (1+z)^n & z>1\ \&\ z\leq 6 \\
%
0 & z>6 
\end{array}    \right., 
\end{equation}
where $m$ and $n$ are free parameters.

\begin{figure}[!ht]
\includegraphics[width=7.5cm]{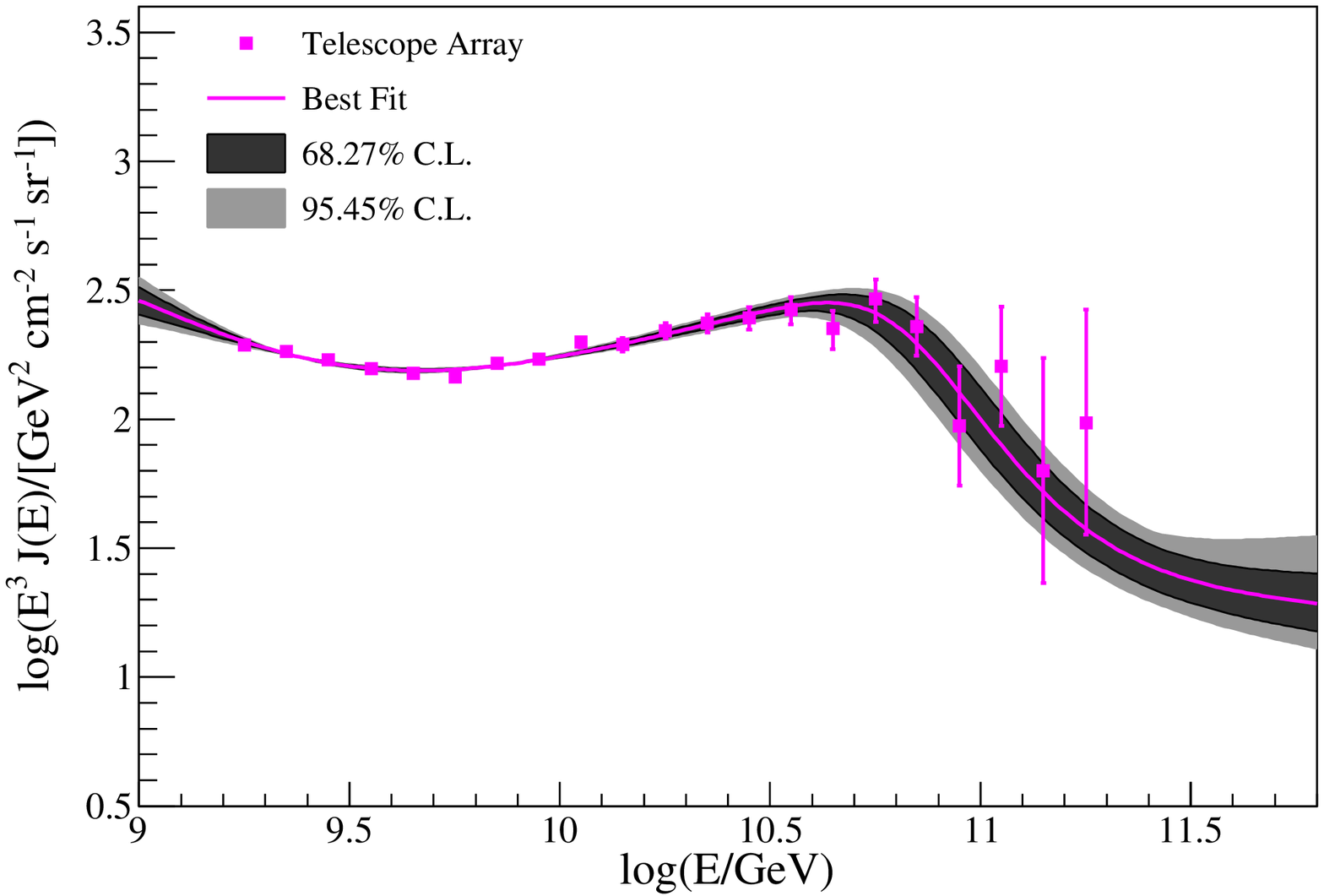}
\includegraphics[width=7.5cm]{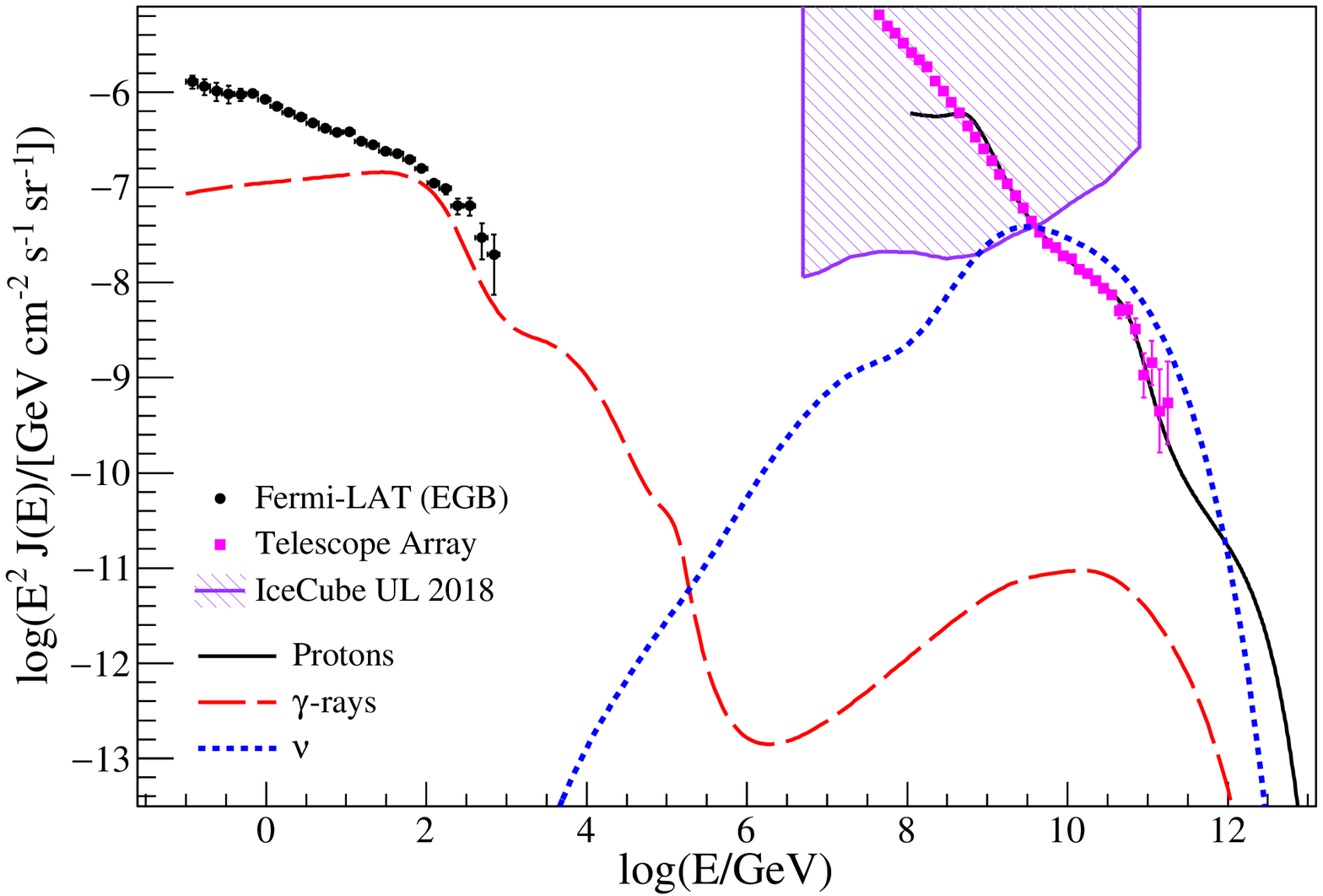}
\caption{Left: Fit of the cosmic ray energy spectrum measured by Telescope Array \cite{TASpec:15}. The solid line 
corresponds to the best fit and the shadowed areas correspond to the 68.27\% and 95.45\% CL regions. Right: Proton, 
gamma-ray, and neutrino spectra corresponding to the best fit of the Telescope Array data. The filled circles 
correspond to the EGB obtained by Fermi-LAT and the shadowed area corresponds to the rejection region at 90\% CL 
for the upper limit on the neutrino flux obtained by IceCube. The EBL model of Ref.~\cite{Kneiske:04} is considered 
and $E_{cut}=10^{21}$ eV.\label{CRFlux}}
\end{figure}

The proton, gamma-ray and neutrino spectra at Earth are calculated by using the TransportCR program \cite{TCR:15}, which 
has been developed to solve numerically the transport equations that govern the propagation of the cosmic rays in the 
intergalactic medium. The Telescope Array flux \cite{TASpec:15} is fitted above $10^{18.2}$ eV in order to avoid the 
effects introduced by a non null intergalactic magnetic field. The parameters $\alpha$ and $m$ are free during the 
minimization procedure. The normalization of the flux and the $\delta_E$ parameter, which takes into account the possible 
systematic uncertainties in the primary energy determination (see Ref.~\cite{Supanitsky:16} for details), are taken as 
nuisance parameters and the profile likelihood technique is used to get rid of them. The contribution of the sources with 
$z>1$ is negligible for $E \geq 10^{18.2}$ eV. Therefore, $\alpha$ and $m$ are determined by fitting the cosmic ray energy 
spectrum above $10^{18.2}$ eV. The left panel of Fig.~\ref{CRFlux} shows the fit of the cosmic ray flux for the case in 
which the cosmic ray sources inject protons below $z=1$ only ($n \rightarrow -\infty$). The EBL model of 
Ref.~\cite{Kneiske:04} is considered in this calculation and the cutoff energy is $E_{cut}= 10^{21}$ eV. The best fit is 
obtained for $\alpha = 2.16$, $m = 6.78$, and $\delta_E = -0.096$. The right panel of the figure shows the best fit of the 
cosmic ray energy spectrum and also the corresponding gamma-ray and neutrino spectra. It can be seen that, unlike the result 
obtained in Ref.~\cite{Supanitsky:16}, the neutrino flux is in tension with the IceCube upper limit. This is due to the 
fact that the upper limit to the neutrino flux recently obtained by IceCube is more restrictive than the one considered in 
Ref.~\cite{Supanitsky:16}.

\begin{figure}[!ht]
\includegraphics[width=7.5cm]{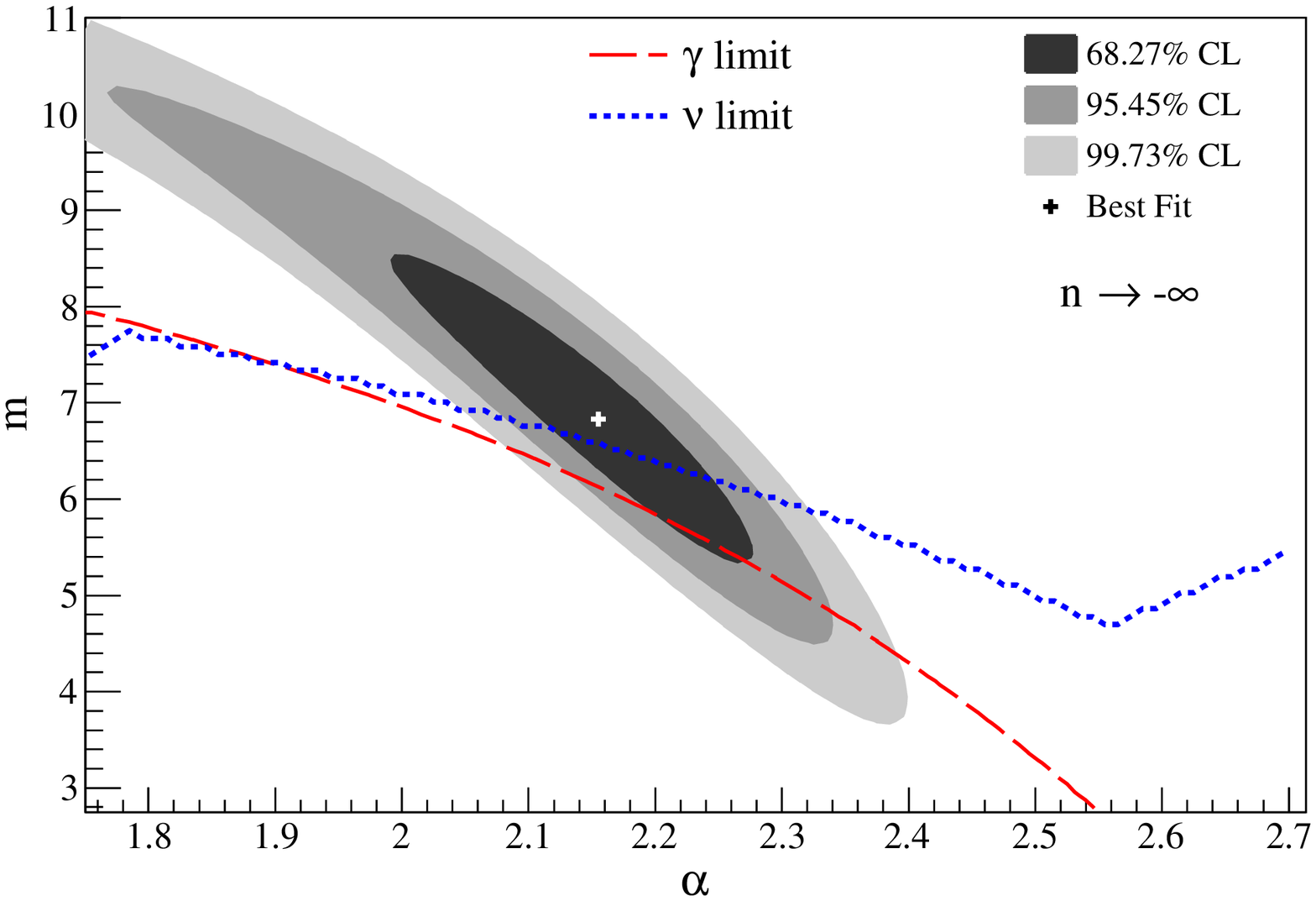}
\includegraphics[width=7.5cm]{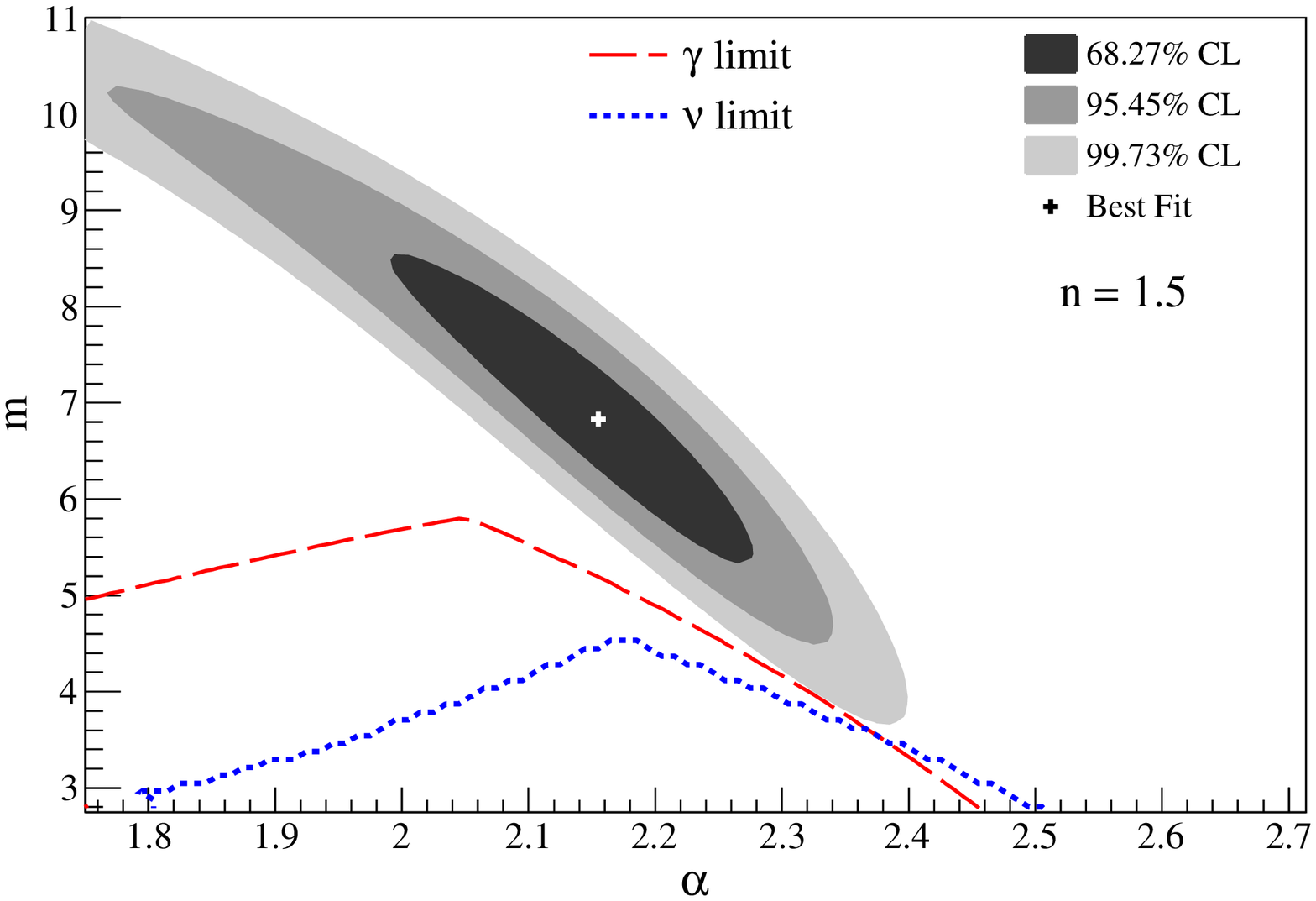}
\caption{Left: Best fit and confidence regions for $n \rightarrow -\infty$. Right: Best fit and confidence regions for 
$n=1.5$. The allowed regions corresponding to gamma-ray and neutrino observations are below the dashed and dotted curves, respectively. The EBL model of Ref.~\cite{Kneiske:04} is considered and $E_{cut}=10^{21}$ eV. \label{FitPar}}
\end{figure}

The left panel of Fig.~\ref{FitPar} shows the best fit and the 68.27\%, 95.45\%, and 99.73\% CL regions of the fit
for $n \rightarrow -\infty$. Also shown are the allowed regions corresponding to the gamma-ray upper limit on the 
integrated gamma-ray flux $I_\gamma^{UL}$, which is given by $I_\gamma(m,\alpha,n) \leq I_\gamma^{UL}$, and form 
the upper limit on the neutrino flux obtained by IceCube (see Fig.~\ref{Nu}), which is given by 
the condition $J_\nu(E,m,\alpha,n) \leq J_\nu^{UL}(E)$. Note that the allowed regions corresponding to the gamma-ray 
and neutrino observations are obtained from the upper limits at 90\% CL. It can be seen from the plot that the best 
fit is in tension with both upper limits. In this case the upper limit obtained from gamma-ray data is more restrictive 
than the one coming from the neutrino data. The right panel of the figure shows the result obtained in the same 
conditions as the ones corresponding to the left panel but for $n=1.5$. It can be seen that the best fit and the 
allowed regions are, as expected, unaltered, but in this case even the region corresponding to 99.73\% CL is in 
tension with the gamma-ray and neutrino upper limits. Models with parameter $n$ larger than $1.5$ have associated 
a larger production of secondary gamma rays and neutrinos. Therefore, for any model with $n$ larger than 1.5 the 
region corresponding to 99.73\% CL of the fit is also in tension with the gamma-ray and neutrino upper limits.

From Fig.~\ref{FitPar} it can be seen that for the case in which the UHECRs are generated in the redshift range from $z=0$ to 
$z=1$, the gamma-ray upper limit imposes more restrictive conditions than the ones corresponding to the neutrino upper limit. 
However, when the sources produce UHECRs beyond $z=1$, the restrictions obtained from gamma-ray and neutrino observations are
complementary, i.e., by using both the neutrino and gamma-ray upper limits, it is possible to enlarge the rejection region. 

Similar results are obtained considereing the EBL model of Ref.~\cite{Inoue:12}. In particular in this case, also for 
$n\geq 1.5$ the neutrino upper limit is in tension with the $(\alpha,m)$ values of the 99.73\% CL region of the fit. 
Considering a cutoff energy of $E_{cut}=10^{19.7}$ eV, as in Ref.~\cite{Supanitsky:16}, and the EBL model of 
Ref.~\cite{Kneiske:04}, it can be seen that for $n\rightarrow-\infty$ the parameters $\alpha$ and $m$ corresponding to the 
best fit are marginally compatible with both the neutrino and gamma-ray upper limits. In this case models with $n\leq 1$
are compatible with the neutrino upper limit.  

As mentioned before, the production of gamma-rays and neutrinos is smaller in models that include heavier nuclei.
In general, mixed composition models that fit the UHECR spectrum and the composition profile measured by Auger are 
compatible with the constraints imposed by current gamma-ray and neutrino observations 
\cite{Unger:15,Vliet:17,Globus:17,Supanitsky:18,Fang:18,Batista:18,Rodrigues:18,Biehl:18}. Only models with a very 
strong evolution are excluded by current observations \cite{Globus:17}. Moreover, it has been shown in 
Ref.~\cite{Batista:18} that the neutrino flux predicted in these types of models can be detected by next generation
neutrino observatories.

\section{Conclusions}

Gamma-ray and neutrino observations impose important restrictions on models of ultrahigh energy cosmic rays. In 
particular, these types of constraints can be used to test the presence of heavier nuclei in the flux in an independent 
manner from the high energy hadronic interaction models, which are used to infer the composition of the cosmic rays
and are subject to relevant systematic uncertainties.

It has been shown that proton models of ultrahigh energy cosmic rays with a strong source density evolution are disfavor
by present gamma-ray and neutrino observations. However, there are proton models that are still compatible with these 
observations. Mixed composition models that fit both the spectrum and the composition profile obtained by Auger are 
less constrained by the gamma-ray and neutrino observations. Only models with a very strong source density evolution are
disfavored by current observation. It is worth mentioning that the neutrino flux associated with these mixed composition 
models will be within reach of next generation neutrino observatories.

\end{document}